\documentclass[journal=jctc,layout=twocolumn,manuscript=article]{achemso}
\SectionNumbersOn 

\usepackage[T1]{fontenc} 		
\usepackage{amsmath,amssymb}
\DeclareMathOperator{\Tr}{Tr}
\usepackage{bm}
\usepackage{changepage}
\usepackage[utf8x]{inputenc}
\usepackage{nameref,hyperref}
\usepackage{microtype}
\DisableLigatures[f]{encoding = *, family = * }
\usepackage[usenames,dvipsnames,svgnames,table]{xcolor}
\usepackage{array}
\usepackage{url}
\usepackage{dcolumn}
\usepackage{graphicx}
\usepackage{multirow}
\usepackage{lineno}
\usepackage{txfonts}
\usepackage{booktabs}
\usepackage{subcaption}
\usepackage{adjustbox}
\usepackage{wrapfig}
\usepackage{siunitx}

\newcommand{\fig}[1]{Figure~\ref{#1}}

\newcommand{\eq}[1]{Equation~\ref{#1}}

\newcolumntype{d}[1]{D{.}{.}{#1}}

\title{Size-and-Shape Space Gaussian Mixture Models for Structural Clustering of Molecular Dynamics Trajectories.}

\author{Heidi Klem}
\affiliation{Department of Chemistry, Colorado State University, Fort Collins, CO 80523}
\author{Glen M. Hocky}
\affiliation{Department of Chemistry, New York University, New York, NY 10003}
\author{Martin McCullagh}
\affiliation{Department of Chemistry, Oklahoma State University, Stillwater, OK 74078}
\email{martin.mccullagh@okstate.edu}

\date{\today}

\begin{document}

\begin{abstract}
Determining the optimal number and identity of structural clusters from an ensemble of molecular configurations continues to be a challenge.  Recent structural clustering methods have focused on the use of internal coordinates due to the innate rotational and translational invariance of these features.  The vast number of possible internal coordinates necessitates a feature space supervision step to make clustering tractable, but yields a protocol that can be system type specific.  Particle positions offer an appealing alternative to internal coordinates, but suffer from a lack of rotational and translational invariance, as well as a perceived insensitivity to regions of structural dissimilarity.
Here, we present a method, denoted shape-GMM, that overcomes the shortcomings of particle positions using a weighted maximum likelihood (ML) alignment procedure. This alignment strategy is then built into an expectation maximization Gaussian mixture model (GMM) procedure to capture metastable states in the free energy landscape.  The resulting algorithm distinguishes between a variety of different structures, including those indistinguishable by RMSD and pair-wise distances, as demonstrated on several model systems.  Shape-GMM results on an extensive simulation of the the fast-folding HP35 Nle/Nle mutant protein support a 4-state folding/unfolding mechanism which is consistent with previous experimental results and provides kinetic detail comparable to previous state of the art clustering approaches, as measured by the VAMP-2 score.  Currently, training of shape-GMMs is recommended for systems (or subsystems) that can be represented by $\lesssim$ 200 particles and  $\lesssim$ 100K configurations to estimate high-dimensional covariance matrices and balance computational expense.  Once a shape-GMM is trained, it can be used to predict the cluster identities of millions of configurations.

\end{abstract}

\section{Introduction}
Structural clustering of molecular dynamics (MD) simulation data of macromolecules is often necessary to draw physical conclusions about molecular mechanisms from large amounts of simulation data.
Consider, for example, the challenge of protein folding.  At or near its folding temperature, a protein will populate both folded and unfolded states.  Molecular simulations  used to probe this process yield a trajectory of atomic positions that approximate the equilibrium conformational ensemble of the molecule. To study the mechanisms of folding and unfolding, it is useful to discretize the protein configurations that are observed within this trajectory.  Which configurations are folded, which are unfolded? Are there multiple folded and/or unfolded states? Are there metastable states along the folding pathway(s)?

To address these, and other similar questions, it is helpful to employ automated structural clustering methods on the trajectory data.  In time-continuous simulations, kinetic clustering, such as spectral clustering of the transition matrix\cite{Deuflhard2000,Deuflhard2005,Kannan2020}, can be employed (see Refs ~\citenum{Keller2010,Peng2018,Glielmo2021} for comparisons between kinetic and structural clustering).  Alternatively, clustering solely on the basis of structural data (structural clustering) can be applied to either time-continuous or disjoint simulation data. Here, we focus on developing a robust structural clustering procedure that can be readily applied to a broad range of systems and depends on only a small number of user-specified, readily understandable, parameter choices.  Structural clustering requires the choice of both a clustering algorithm and the features on which to perform the clustering.  We start our discussion with the choice of features, which can be separated into two categories: internal coordinates and particle positions.

Internal coordinates of a macromolecule are an appealing choice of features because they are invariant to both translation and rotation.  These coordinates include distances (two-body terms), angles (three-body terms), dihedral angles (four-body terms), and higher order terms not often considered for clustering purposes.  The inclusion of even all two-body terms yields a feature space that drastically overdetermines the system and hinders the ability to cluster in this space. Possible solutions to the over determination problem include the use of \textit{a priori} knowledge of the system to develop a hypothesis-driven selection of key internal coordinates, segmentation of the trajectory\cite{Damjanovic2021}, combining both kinetic and structural information\cite{Cocina2020}, and unsupervised dimensionality reduction techniques.  We note that applying dimensionality reduction to internal coordinates is non-trivial as projection errors can misconstrue data\cite{Sittel2017}.  Recent efforts, even methods that combine kinetic and structural information, have focused on choosing a subset of features, e.g. backbone dihedral angles of proteins, that are able to distinguish between perceived important structural states of the macromolecule of interest.\cite{Altis2008,Sittel2018}  This supervision step, however, yields a protocol that is specific to the system/problem of interest. Thus, it remains a challenge to use internal coordinates in a general structural clustering workflow.

Particle positions offer an appealing alternative choice of features because they do not drastically overdetermine a system.  A major limitation of particle positions is that they are defined in the lab frame and thus are not invariant to translation and rotation.\cite{Sittel2018}
Previously, structural alignment based on root-mean-squared displacement (RMSD) has been used to overcome rotational and translational invariance, and subsequent RMSD between structures has been used as a distance metric in clustering algorithms\cite{Daura1999,Keller2010}.  This has led to the conclusion that particle positions have a second disadvantage: RMSD between structures is a poor estimate of similarity for systems in which there is regional heterogeneity in particle fluctuations \cite{Daura1999,Keller2010}.  Below, we will demonstrate that this conclusion is due to the assumption, implicit in RMSD alignment, that particles vary in independent, equivalent, spherical distributions.

Numerous clustering algorithms have been applied to structural data from MD simulations but the connection between the resulting clusters and the underlying assumptions of the algorithm are often obscure. Recent efforts have focused on the use of non-hierarchical algorithms applied to a reduced dimensional subspace of the original features.  These algorithms include k-means\cite{Shao2007,Bowman2009,Jain2012},  Gaussian mixture models (GMMs)\cite{Westerlund2019}, and density-based clustering schemes \cite{Sittel2016,Rodriguez2014,Melvin2016,Trager2021}.  Unsupervised dimensionality reduction has been achieved using principal component analysis (PCA), time-lagged independent component analysis (TICA), sketch map\cite{Ceriotti2011}, UMAP\cite{Trozzi2021}, or a variety of other methods\cite{Glielmo2021}. The combination of internal coordinates (e.g. distances, angles and/or dihedrals) and dimensionality reduction schemes makes it challenging to predict the expected form of the distribution in these spaces.  This has led to the popularity of density-based clustering schemes.  The downside of these schemes are two-fold: first, there is a perception that they can only be applied in fairly low-dimensional spaces and second, there is not a strong theoretical connection between the clustering algorithm and the physics governing the resulting conformational clusters.

In this paper, we present shape-GMM, a method to cluster macromolecule trajectory data directly using particle positions.
Previous shortcomings of RMSD clustering are overcome using a more general Mahalanobis distance and a weighted maximum likelihood (ML) alignment procedure.\cite{Theobald2006} The shape-GMM algorithm incorporates this rigorous alignment strategy into an expectation maximization (EM) GMM procedure. The GMM approach is conceptually appealing because it resembles a first-order approximation to the expected probability density of particle positions.  The ability of shape-GMM to correctly cluster molecular structures is first assessed using a set of elastic network model systems with known "ground truths".  Subsequently, the real-world applicability of this method is demonstrated on a challenging, well-studied protein folding system, and the results are shown to be consistent with previous experimental and theoretical studies.  Combined, this work establishes shape-GMM as an appealing clustering approach to a broad range of macromolecular structures.

\section{Theory and Methods}
\subsection{Molecular Size-and-Shape}
We consider a MD simulation of a macromolecule in solution.  From a single frame, the macromolecule is represented by $N$ atoms considered important (e.g. all protein, all heavy atoms, just $C_\alpha$ atoms, etc.) or an $N$ particle coarse-grained mapping (e.g. the centers of mass of each protein residue). These features are a point in $\mathbb{R}^{3N}$ encoded by a matrix, $\bm{x}$, of order $N\times3$. The Hamiltonian for any system considered here is independent of the lab-frame, and so the features of interest, $[\bm{x}_i]$, are the orbit of all possible rigid-body transformations of $\bm{x}_i$.  This is written as an equivalence class, 
\begin{equation}
[\bm{x}_i] = \{\bm{x}_i\bm{R}_i + \bm{1}_N\vec{\xi}_i^T : \vec{\xi}_i \in \mathbb{R}^3, \bm{R}_i \in \text{SO}(3)  \},
\end{equation}
where $\vec{\xi}_i$ is a translation in $\mathbb{R}^3$, $\bm{R}_i$ is a rotation $\mathbb{R}^3\rightarrow\mathbb{R}^3$, and $\bm{1}_N$ is the $N\times1$ vector of ones. The features, $[\bm{x}_i]$ exist as a point in size-and-shape space\cite{Dryden1998}. Size-and-shape space has dimensions $3N-6$ and is defined as $S\Sigma_N^3 = \mathbb{R}^{3N}/G$ where $G = \mathbb{R}^3\times \text{SO}(3)$ is the group of all rigid-body transformations for each frame with elements $g=( \vec{\xi},\bm{R})$. 

For a trajectory consisting of $M$ frames or $M$~molecular configurations, the set of recorded points $\{\bm{x}_1,\ldots, \bm{x}_M\}$ is effectively regarded as a set of orbits
\begin{equation}
\{[\bm{x}_1],\ldots, [\bm{x}_M] \} = \{g_i^{-1} \bm{x}_i  : g_i \in G \}.
\end{equation}
The group $G$ is said to act freely, or $G^M$ acts component-wise, one group element per frame. 

\subsection{Gaussian Distribution of Positions}

If we consider the macromolecule to be in a single free energy minimum, the expected first-order approximation to the probability density is Gaussian in $\mathbb{R}^{3N}$.
The normalized multivariate Gaussian distribution is given as
\begin{equation}
\label{mvGaussian}
N(\bm{x}_i  \mid \bm{\mu}, \bm{\Sigma}) = \frac{ \exp\left[ -\frac{1}{2}(g_i^{-1} \bm{x}_i-\bm{\mu})^T\bm{\Sigma}^{-1}(g_i^{-1} \bm{x}_i-\bm{\mu})\right] } {\sqrt{(2\pi)^{(3N)}\det{\bm{\Sigma}}}},
\end{equation} 
where $\bm{\mu}$ and $\bm{\Sigma}$ are the mean and covariance ($3N\times3N$), respectively, and the multiplication inside the exponent requires a flattening of the $(g_i^{-1} \bm{x}_i-\bm{\mu})$ matrix.  The estimation of well-defined average and covariance matrices from an MD trajectory requires determining the appropriate set of rigid body transformations, $(g_1,\ldots, g_M)$, for which we define a maximum likelihood (ML) procedure. 

\subsection{Maximum Likelihood Alignment}
\label{sec:ML_alignment}

The following ML alignment is adapted from Theobald and Wuttke\cite{Theobald2008} in two ways.  First, we generalize the allowed form of the covariance and, second, we do not enforce an inverse gamma distribution of the eigenvalues of the covariance because we expect sufficient sampling of this matrix from the MD trajectory. The log likelihood of a trajectory alignment is given as
\begin{equation}
    \label{ln_l_alignment}
    \ln(L) = \sum_{i=1}^M \ln(N(\bm{x}_i  \mid \bm{\mu}, \bm{\Sigma})).
\end{equation}
There are two distinct components for the parameters in a single trajectory ML alignment:
\begin{enumerate}
\item the alignment parameters or group elements $(g_1,\ldots, g_M)$; and
\item the mean configuration~$\bm{\mu}$, and the covariance matrix~$\bm{\Sigma}$.
\end{enumerate}
The translation alignment parameters are easily resolved by removing the center of geometry from each configuration. Methods to determine optimal rotation matrices have only been identified for covariances of the form $\bm{\Sigma}\propto \bm{I}_{3N}$\cite{Kabsch1976,Horn1987} or $\bm{\Sigma} = \bm{\Sigma}_N\otimes \bm{I}_3$\cite{Goodall1991,Theobald2006}, where $\otimes$ is a Kronecker product and $\bm{\Sigma}_N$ is the $N\times N$ covariance matrix. We note that a covariance of the form  $\bm{\Sigma} \propto \bm{I}_{3N}$ assumes that particles oscillate in uncoupled, identical, spherical distributions.  Here, we will employ the forms $\bm{\Sigma} = \bm{\Sigma}_N\otimes \bm{I}_3$ (``weighted'') and  $\bm{\Sigma}\propto \bm{I}_{3N}$ (``uniform'').

The problem of finding the parameter values that maximize the likelihood can be solved by iterating between two steps until the log likelihood (\eq{ln_l_alignment}) has converged within some threshold.
\begin{enumerate}
    \item Given $g_1,\ldots, g_m$, calculate $\bm{\mu}, \bm{\Sigma}$ by
\begin{eqnarray}
\bm{\hat x}_i &=& g_i^{-1} \bm{x}_i, \\
\bm{\hat\mu} &=& \sum \frac{\bm{\hat x}_i}{M}, \\
\bm{\hat\Sigma}_N &=& \sum \frac{(\bm{\hat x_i} - \bm{\hat\mu}) (\bm{\hat x_i} - \bm{\hat\mu})^T}{3M} 
\end{eqnarray}
For the sub-model in which $\bm{\Sigma} = \sigma^2 \bm{I}_N$, it suffices to take $\hat\sigma^2 = \Tr(\bm{\hat\Sigma}_N)/N$.
\item Given $\bm{\hat\mu}, \bm{\hat\Sigma}$ and an appropriate inverse $\bm{W} = \bm{\Sigma}^{-1}$, estimate the alignments $g_1,\ldots, g_m$ by minimizing the Mahalanobis distance
\begin{equation}
\label{eq:distance}
||g_i^{-1}\bm{x}_i - \bm{\hat\mu}||^2 = \Tr\bigl((g_i^{-1}\bm{x}_i - \mu)^T W (g_i^{-1}\bm{x}_i -  \bm{\mu})\bigr)
\end{equation}
with respect to $g_i$.
\end{enumerate}

The second step is the generalized Procrustes problem, which can be solved either by using the singular-value decomposition\cite{Kabsch1976,Goodall1991,Theobald2006,Theobald2008} or by quaternion methods\cite{Horn1987,Theobald2005,Liu2010}.  For the case that $\bm{\Sigma}\propto \bm{I}_{3N}$ (``uniform''), \eq{eq:distance} simplifies to the mean squared displacement.  Thus, our uniform covariance model is equivalent to an RMSD alignment and distance metric.

If $\bm{\Sigma}_N$ is unrestricted, the estimate $\bm{\hat\Sigma}_N$ has rank $N-1$ and kernel~$\bm{1}$. In that case, $\bm{W}$~is the spectral inverse which also has kernel~$\bm{1}$. 

\subsection{Gaussian Mixture Model for Size-and-Shape Features}

In a non-harmonic force field description of a globular macromolecule, multiple metastable configurations will be encountered.  It is natural to consider the probability density in this feature space as a mixture of $K$ multivariate Gaussians,\cite{Frauenfelder1988}
\begin{equation}
\label{gmm}
P(\bm{x}_i) = \sum_{j=1}^{K} \phi_j N(\bm{x}_i\mid \bm{\mu}_j, \bm{\Sigma}_j),
\end{equation}
where $\phi_j$ is the weight ($\sum_{j=1}^{K} \phi_j = 1$), $\bm{\mu}_j$ is the mean, and $\bm{\Sigma}_j$ is the covariance of the $j$th Gaussian probability density.   \eq{gmm} will be a good approximation for the probability density near local minima, the preferentially sampled configurations in conventional molecular dynamics. This approximation will fail near transition states although adaptations to this formulation have been developed to account for this\cite{Westerlund2019}. 

\eq{gmm} is known as a Gaussian Mixture Model (GMM), and the goal of determining the optimal $\{\phi_j\}$, $\{\bm{\mu}_j\}$, and $\{\bm{\Sigma}_j\}$ that fit observations has been achieved previously with algorithms including expectation maximization (EM) with a maximum likelihood criterion or Bayesian approaches including variational inference or sampling techniques (e.g Markov Chain Monte Carlo\cite{Fong2012} or Gibbs Sampling\cite{stephensGibbs}).  Any of these approaches can be applied to the trajectory features, $\{[\bm{x}_i]\}$, but care must be taken to correctly account for the feature equivalences.  

\subsubsection{Gaussian Mixture Model: Expectation Maximization}

Within the EM framework, for a fixed number of Gaussians, $K$, the following steps are taken\cite{stephensEM}. Note that in this work two choices are available for step 3, depending on the approximation to the form of $\bm{\Sigma}$ chosen, as described in the previous section.

\begin{enumerate}
\item Provide initial guesses for $\{\hat{\phi}_j\}$, $\{\hat{\bm{\mu}}_j\}$, and $\{\hat{\bm{\Sigma}}_j\}$

\item \textbf{Expectation:} estimate the posterior distribution for latent variable of mixture components, $Z_i \in \{1,...,K\}$:
\begin{equation}
\label{posterior}
\gamma_{Z_i}(j) = \frac{\hat{\phi}_j N(\bm{x_i} \mid \hat{\bm{\mu}}_j, \hat{\bm{\Sigma}}_j)}{\sum_{j=1}^K\hat{\phi}_j N(\bm{x_i}\mid \hat{\bm{\mu}}_j, \hat{\bm{\Sigma}}_j)}
\end{equation}
\item \textbf{Maximization:} update $\{\hat{\phi}_j\}$, $\{\hat{\bm{\mu}}_j\}$, and $\{\hat{\bm{\Sigma}}_j\}$.
\begin{eqnarray}
\label{update_weights}
\hat{\phi}_j  &=& \frac{\sum_{i=1}^N\gamma_{Z_i}(j)}{N}\\
\label{update_avg}
\hat{\bm{\mu}}_j  &=& \frac{\sum_{i=1}^M\gamma_{Z_i}(j)g_{i,j}^{-1}\bm{x}_i}{\sum_{i=1}^M\gamma_{Z_i}(j)}\\
\label{update_var}
\hat{\bm{\Sigma}}_j  &=& \begin{cases}
\frac{\sum_{i=1}^M\gamma_{Z_i}(j)\langle\hat{\sigma}^2\rangle_i}{\sum_{i=1}^M\gamma_{Z_i}(j)}\bm{I}_{3N} & \textrm{uniform}\\
\frac{\sum_{i=1}^M\gamma_{Z_i}(j)\langle\hat{\bm{\Sigma}}_{N}\rangle_i}{\sum_{i=1}^M\gamma_{Z_i}(j)}\otimes\bm{I}_{3} & \textrm{weighted}\\

\end{cases} 
\end{eqnarray}
\item Iterate steps 2-3 until the log likelihood converges within some threshold.  Log likelihood is defined as
\begin{equation}
\ln(L) = \sum_{i=1}^M\ln\left( \sum_{j=1}^K\hat{\phi}_jN(\bm{x}_i \mid \hat{\bm{\mu}}_j, \hat{\bm{\Sigma}}_j)\right)
\end{equation}
\end{enumerate}

\subsubsection{Shape-GMM Procedure}
Shape-GMM adapts the EM algorithm for size-and-shape space by including trajectory alignments in both the Expectation and Maximization steps:
\begin{enumerate}
\item In the Expectation step, the distribution $N(\bm{x_i} \mid \bm{\mu}_j, \bm{\Sigma}_j)$ is determined for each frame after alignment (either uniform or weighted) to the respective average.  This requires $K$ non-iterative trajectory alignments.
\item In the Maximization step, the means and covariances are updated using the ML alignment described in Sec. \ref{sec:ML_alignment}.  This requires $K$ iterative trajectory alignments.  
\end{enumerate}

\subsubsection{Model Initialization}

The EM procedure must be initialized by providing guesses for $\{\hat{\phi}_j\}$, $\{\hat{\bm{\mu}}_j\}$, and $\{\hat{\bm{\Sigma}}_j\}$.  This can be achieved in a variety of ways including dividing the trajectory into $K$ parts, using a k-means algorithm, or randomly selecting initial frames as averages and assigning frames to their nearest (measured by RMSD) cluster center.  All of these procedures are implemented in our code, but we use the random frame initialization for all examples unless explicitly stated otherwise. 

\subsubsection{Assigning Frames to Clusters}

Within the context of clustering from a GMM, a frame (or data point) is assigned to the cluster in which it has the largest likelihood.  This is no different in size-and-shape space.  

\subsubsection{Determining the Number of Clusters}

The number of clusters, $K$, must be specified in order to perform shape-GMM.  The best choice of $K$ will be system specific and is a challenge to determine for any clustering method.  Here, we use the elbow method coupled with cross-validation.  The elbow method is a heuristic method to identify a significant change in slope in the log likelihood (in the case of GMMs) as a function of number of clusters. Cross-validation is a method of applying the trained model on a validation (non-trained) data-set.  This is done as a function of number of clusters and an elbow in this curve is used to pick the number of clusters.   Examples of these plots are given for systems in the Results and Discussion section and SI.  

For cross-validation, we use a simple scheme to break up the data into a training set and a validation set.  The training set is randomly (uniformly) chosen from the entire data set and the remaining data is used for validation.  The number of frames chosen for training is system specific (but at least 1000 for all systems studied here) and is given in the Results and Discussion section for each system.

\subsubsection{Implementation}

The current implementation of shape-GMM is written in python with acceleration/parallelization using the numba package.\cite{Lam2015}  The shape-GMM package is written to mimic the interface of the Scikit-learn Gaussian Mixture package \cite{scikit-learn}.  The package can be easily installed using pip (pip install shapeGMM).  All code, as well as examples of the full analyses performed below are available from \url{https://github.com/mccullaghlab/GMM-Positions}.

\section{Results and Discussion}
We explore the shape-GMM clustering method using a variety of examples. 
The first example we consider is a set of elastic network models (ENMs) designed to probe the ability of both uniform and weighted covariance versions of shape-GMM to distinguish between structures for a known clustering.    
Subsequently, we examine performance on one toy and one real example of models where molecular topology is free to change, to demonstrate the ability of weighted shape-GMM to estimate the number and identity of configurational clusters in the data.  
Full descriptions of MD procedures for each system are provided in the Sec.~\ref{sec:sim_details}.

\subsection{Elastic Network Models}
In order to rigorously compare and contrast clustering protocols, we create amalgamated trajectories of various Elastic Network Models (ENMs).  The force constants and topologies of the ENMs are chosen such that each ENM will populate a single conformation. Specifically, we consider a variety of ENM models with topologies motivated by protein secondary structure elements. 
 To further mimic the nature of combined protein structural elements, we utilize an anisotropic network model (ANM).  The Hamiltonian for our ANM models is given as
\begin{equation}
    H(\bm{x}) = \sum_{i,j>i} k_{ij} (|\vec{r}_{ij}| - |\vec{r}^0_{ij}|)^2,
\end{equation}
where  $|\vec{r}_{ij}| = |\vec{x}_i - \vec{x}_j|$ is the distance between particles $i$ and $j$, $|\vec{r}^0_{ij}|$ is the distance between particles $i$ and $j$ in the reference geometry, and 
\begin{equation}
    k_{ij} = \begin{cases}
    100   & j=i+1 \\
    20   & |\vec{r}^0_{ij}|=1.7 \textrm{\AA} \\
    10 e^{-||\vec{r}^0_{ij}|-1.7|} & \textrm{otherwise},
    \end{cases}
\end{equation}
where $k_{ij}$ has units $\mathrm{kcal \cdot mol^{-1} \cdot \textrm{\AA}^{-2}}$.  The strongest force constant is reserved for primary sequence bonds, ``hydrogen bonds'' are given the second strongest force constant and are set at a separation distance of $1.7$ \AA, and all remaining pairwise interactions are given an exponentially decaying distance dependent force constant.  

A variety of topologies and system sizes of these types of ANMs are used in the subsequent subsections.  Examples of the types of topologies are depicted in \fig{fig:5states_GMMs}(A). We note that because shape-GMM relies only on positions, it applies to these coarse-grained models without modification, whereas many state-of-the art clustering methods designed for protein data either require temporal information or are applied on backbone dihedral angles, and hence we cannot compare to them here. 

\subsubsection{Clustering of 5 ANMs using shape-GMM}
\label{5_enm}

\begin{figure}
    \centering
    \includegraphics[width=0.5\textwidth]{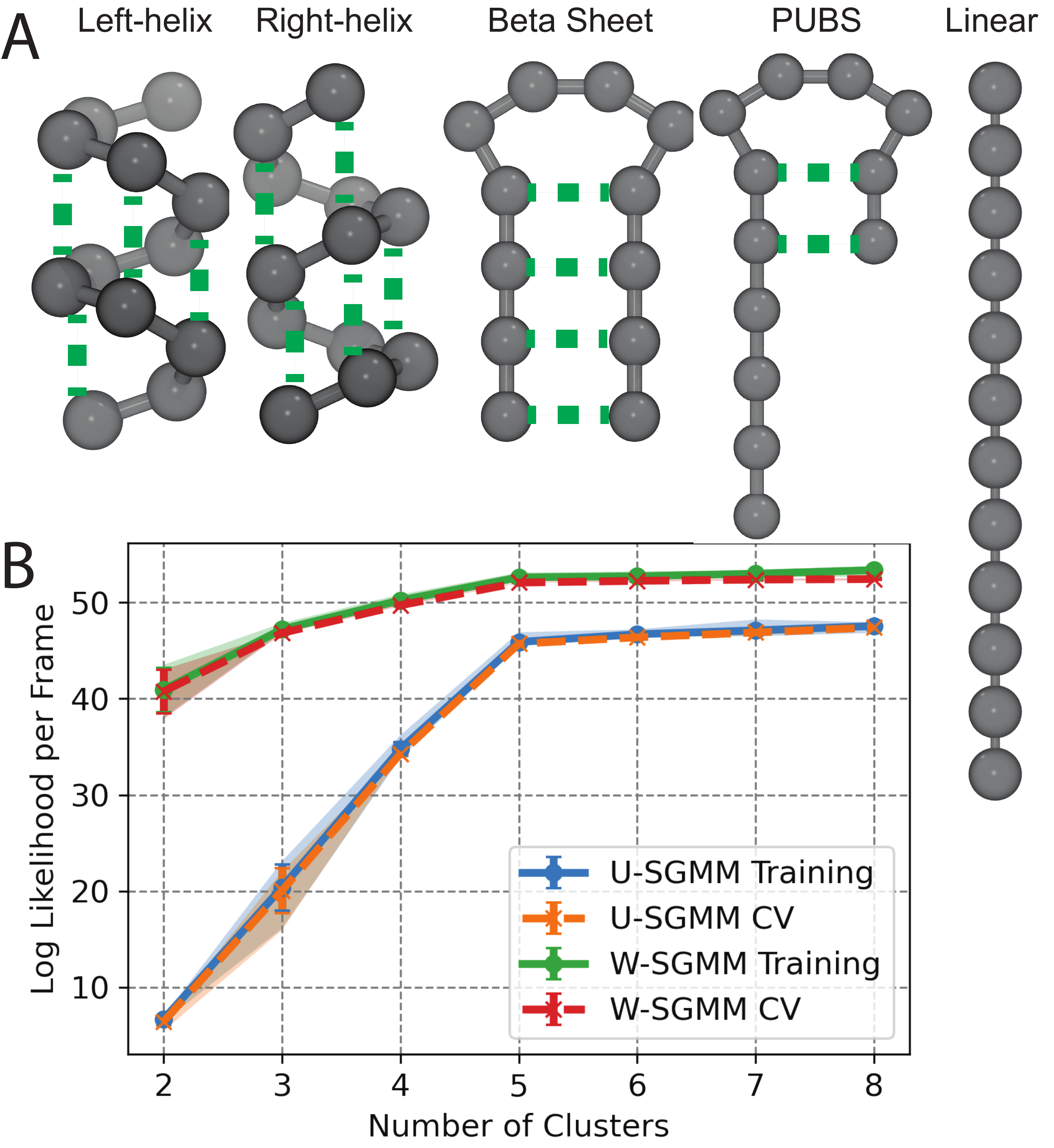}
    \caption{Clustering of an amalgamated trajectory of five anisotropic network models (ANMs) of a 12 bead system.  (A) Schematic of the five topologies considered including left-handed helix, right-handed helix, beta-sheet like, and partially unfolded (PUBS), and linear. (B) Log likelihood per frame as a function of number of clusters for uniform shape-GMM and weighted shape-GMM.  Each model has two corresponding curves: the log likelihood per frame of the training set and the log likelihood per frame from the cross-validation (CV) set.  Error bars are the standard deviation of sampling ten different training sets for each model.  Shaded regions denote the 90\% confidence interval.}
    \label{fig:5states_GMMs}
\end{figure}

To rigorously assess the ability of uniform and weighted shape-GMM to cluster molecular-like configurations, we have constructed a set of five ANMs each with 12 particles depicted in \fig{fig:5states_GMMs}A.  These five structures are motivated by protein secondary structural elements: a left-handed helix, a right-handed helix, a beta sheet-like structure with a four bead hairpin, a partially unfolded beta sheet (PUBS) with both a structured region and an unstructured region, and a linear chain that might represent an unfolded or an unstructured region of a protein.  Equal length simulations (10000 frames each) of each model were run and a single trajectory was created by concatenation.  Clusterings can be compared to the ground truth (denoted by which ANM simulation the frame comes from) using correctly clustered frame pairs to avoid cluster index invariance.  

Shape-GMM is able to correctly identify five clusters in the amalgamated trajectory of the five ANMs depicted in \fig{fig:5states_GMMs}A.  To demonstrate this, we trained both uniform and weighted shape-GMMs for numbers of clusters varying from 2 to 8 on 2000 randomly selected frames; this process was repeated on five different training sets to demonstrate repeatability (and to estimate error) and the cross-validation set was the remaining 48K frames for each set. The results for uniform and weighted shape-GMM are comparable so we will only discuss the uniform results for clarity.  The resulting relative log likelihood per frame of the uniform shape-GMM is plotted as a function of number of clusters as the blue (training) and orange (cross-validation) curves in \fig{fig:5states_GMMs}B.  The log likelihoods of both the training set and the cross-validation set rapidly increase from 2 to 5 clusters at which point the values plateau. The plateau in log likelihood at 5 clusters indicates that increasing the number of clusters does not improve the fit of the model to the data.  Additionally, a lack of measurable deviation between the training set and the cross-validation set indicates no over or under fitting of the model.  Thus, uniform shape-GMM on this data suggests choosing five clusters.  Both uniform and weighted five cluster shape-GMMs yield 100\% agreement with the ground truth on the cluster assignments.

\subsubsection{Clustering of 5 ANMs using Internal Coordinates}

Internal coordinates can be used to cluster these same five ANM structures, but an accurate clustering requires a combination of pairwise distances and backbone dihedrals.  Internal coordinates do not comprise a vector space and thus the distributions in these spaces will not be strictly Gaussian.  Thus, we employ a  density-based clustering algorithm, CLoNe, that has been recently applied to molecular dynamics data\cite{Trager2021}.  CLoNe has a single free parameter, denoted pdc, that dictates the maximum distance to consider grouping points together.  Altering this parameter can change the number and identity of clusters in the data.  For all feature spaces investigated, we scanned this parameter and chose a value that optimized the clustering overlap with the ground truth.  

CLoNe on pairwise positions or backbone dihedrals is insufficient to adequately cluster the amalgamated 5 ANM trajectory.  In both cases, CLoNe predicts four major clusters (some frames are assigned as noise).  CLoNe on all pairwise distances (pdc=6; 66 total distances) unsurprisingly groups the two helix structures into one cluster as pairwise distances take on the same values in the two clusters.  This yields a clustering overlap with the ground truth of 91.3 \%.  CLoNe on backbone dihedrals (pdc=5.7; 18 values encoded by sin and cos of the 9 dihedrals) divides the linear structure into two clusters due to the ability of each dihedral to take on a wide array of values.  The resulting overlap with ground truth is 88.6 \%.  CLoNe on a feature space composed of backbone dihedrals and pairwise distances (pdc=4.1; 84 total features) predicts the proper five clusters and yields an overlap with ground truth of 99.9\%.  

While CLoNe on the combination of internal coordinates is able to properly cluster the five ANM models, this agreement requires two significant supervision steps.  First is the choice of specific internal coordinates.  It is common to choose one or the other of these coordinates but there are some cases in which both are important.  This could be a situation like the combined dissociation of tertiary and secondary structural elements during protein unfolding.  The second supervision step is the choice of the pdc parameter.  Here, we chose a pdc parameter that yielded an improved agreement with ground truth.  Clearly, that is not feasible in all situations.

\subsubsection{Uniform versus Weighted Alignment in Shape-GMM}
\label{sec:uniform_v_weighted}
\begin{figure}
    \centering
    \includegraphics[width=\columnwidth]{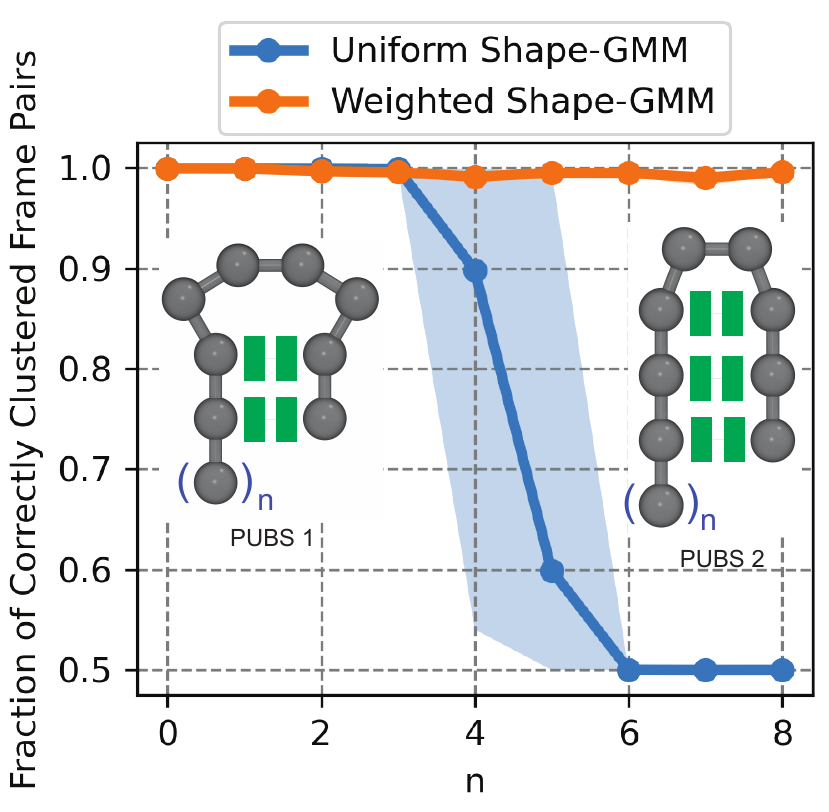}
    \caption{Clustering of two anisotropic network models (ANMs) with varying length of an unfolded region.  Fraction of correctly clustered pairs of frames as a function of the size of unfolded region for uniform shape-GMM and weighted shape-GMM. Shaded regions denote the 90\% confidence interval obtained by clustering with 10 different training sets. Schematics of the two topologies of ANMs used in the clustering are provided as insets.}
    \label{fig:2states}
\end{figure}

Uniform shape-GMM is unable able to distinguish between subtle structural differences in systems with enhanced flexibility throughout different regions of the molecule.  This is because uniform shape-GMM relies on a RMSD-based alignment that assumes particles fluctuate independently in equivalent, spherical distributions.  To demonstrate this failing, we look at two partially unfolded beta sheet ANM models, named PUBS 1 and PUBS 2 and depicted in the insets in \fig{fig:2states}.  The only difference between the two structures is that two beads in the four bead loop region of PUBS 1 are moved into ``hydrogen bonding'' distance in PUBS 2.  The size of the ``unfolded'' region is varied from zero to eight beads in each system to assess the impact of changing the size of the highly varying region.  Simulations of each of these were performed independently and an amalgamated trajectory was created with 5000 frames from each simulation.  

Uniform shape-GMM is unable to distinguish the two partially unfolded beta sheet structures when the unfolded region becomes comparable in size to the folded region.  To quantify this, we performed uniform shape-GMM on combined trajectories of PUBS 1 and PUBS 2 for varying sizes of the unfolded region (quantified by $n$ in \fig{fig:2states}).  The number of clusters was fixed at 2 and the model was initiated with the correct clustering (to avoid convergence to local maxima).  The fraction of correctly clustered pairs of frames from the output clustering was then calculated as a function of unfolded region size (\fig{fig:2states}).  The uniform shape-GMM curve (blue), correctly clusters almost all frame pairs for unfolded region sizes from 0 to 3 beads.  For $3<n<7$, the uniform shape-GMM overlap with the ground truth clustering rapidly decays to only getting 50\% of the frame pairs correct, the minimum possible given that each frame must be clustered into one of two clusters.  Thus, uniform shape-GMM is unable to distinguish between configurations PUBS 1 and PUBS 2 for unfolded regions of five or more beads, situations in which the folded region and unfolded region are comparable in size.  This result is consistent with previous findings based on pairwise RMSD clustering. 

Weighted shape-GMM can distinguish between the two partially unfolded beta sheet structures for all unfolded regions sampled.  Again, this is quantified by the clustering overlap with the ground truth as a function of unfolded region size in \fig{fig:2states}B.  The weighted shape-GMM curve, depicted in orange, demonstrates a fraction of correctly clustered frame pairs near $1.0$ for all sampled unfolded region sizes.  This result demonstrates that the covariance in the weighted model is a better match to the actual covariance from the PUBS 1 and PUBS 2 ANMs than the uniform covariance.  Additionally, this result indicates that the shape-GMM procedure can correctly capture subtly different behavior in heterogeneously fluctuating molecules such as the ones depicted here. This result exhibits how weighted shape-GMM overcomes a previous challenge that limited the applications of particle positions as features for clustering.

\subsection{Beaded Helix Transitions}

\begin{figure}
    \centering
    \includegraphics[width=\columnwidth]{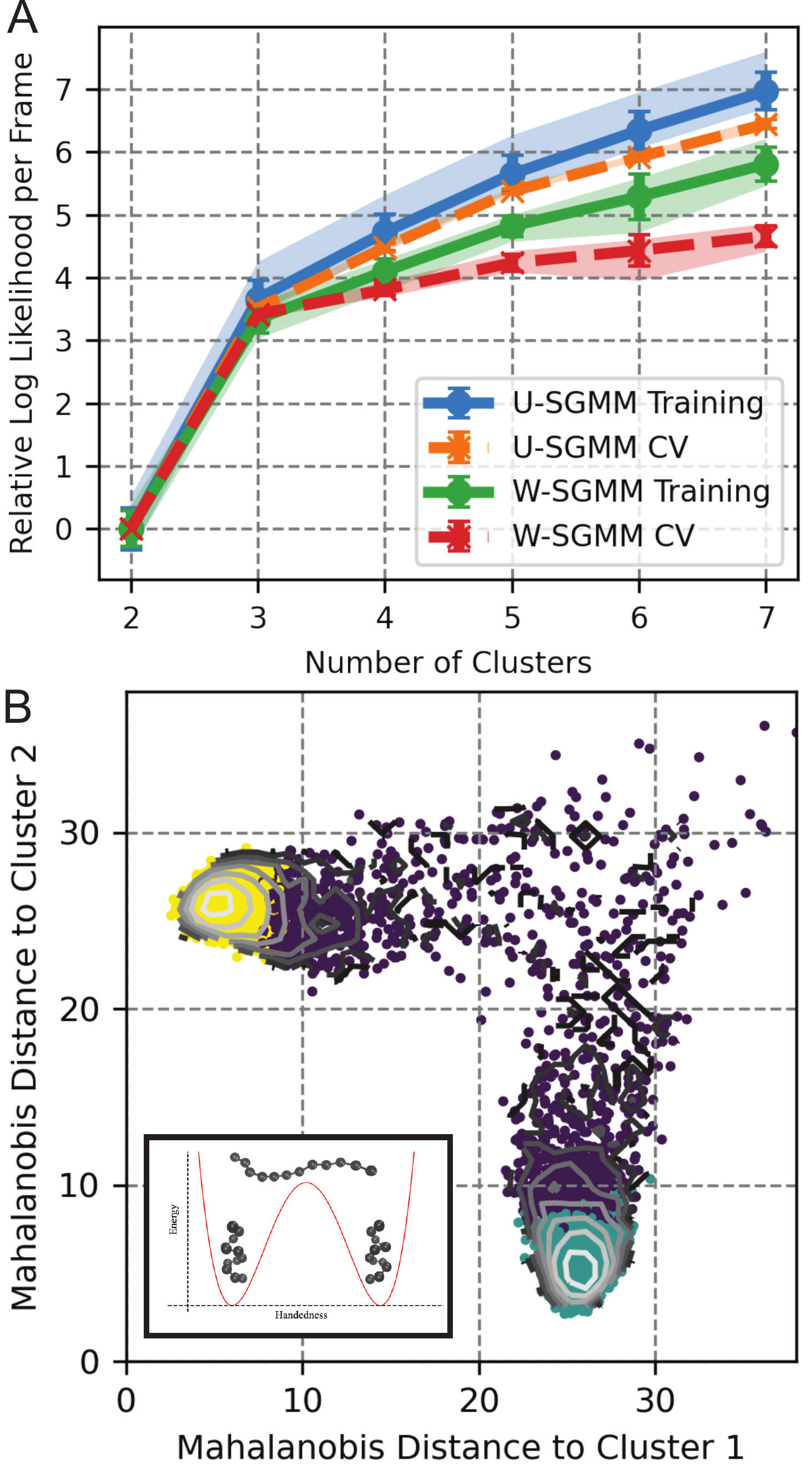}
    \caption{Clustering of a 12-bead helix transition trajectory. A) The Log likelihood as a function of number of clusters for uniform shape-GMM and weighted-shape-GMM. Error bars denote standard deviation and shaded regions denote the 90\% confidence interval obtained from clustering using five different training sets. B) Clustering results for 3 clusters from W-SGMM on 2D free energy surface of weighted Mahalanobis distance from two cluster centers (inset: schematic of potential energy diagram).}
    \label{fig:beaded_helix}
\end{figure}

To assess the ability of the shape-GMM to cluster a non-harmonic system, we consider a 12-bead helix system with harmonic bonds along the backbone and Lennard-Jones attractions between every fifth bead described in Ref.~ \citenum{Hartmann2020}.
Depending on the strength of attraction between every $i,i+4$ bead (or nearly equivalently, the temperature), this model can be trapped in its starting helix or make transitions between left- and right-handed helices. A schematic of this process is depicted in the inset of \fig{fig:beaded_helix}B. 
We choose an attractive strength of $\epsilon=6k_B T$, in which case the helices are stabilized but some rare transitions can occur.
We performed both uniform and weighted shape-GMM on a 10000 frame trajectory to assess the ability of these models to cluster the data.

Both uniform and weighted shape-GMM support the choice of three clusters to represent the simulation data. A plot of the log likelihood per frame as function of number of clusters is depicted in\fig{fig:beaded_helix}A for both uniform and weighted.  In this scan, five randomly selected training sets of 1000 frames are chosen.  Uniform and weighted shape-GMMs are fit with the training set and then cross-validated on the remaining 9000 frames. Both uniform and weighted models have a significant change in slope at 3 clusters for both training and cross-validation curves.  Additionally, the cross-validation curve for weighted shape-GMM (red) starts to deviate significantly from the training curve (green) as the number of clusters is increased above 3.  This behavior indicates an overfitting of the weighted shape-GMM above three clusters.  The overfitting behavior is less significant for the uniform shape-GMM but is still present and, as seen in the previous section, a weighted shape-GMM is able to differentiate between heterogeneously fluctuating systems more readily than a uniform shape-GMM.

Weighted shape-GMM identifies three states corresponding to the left-handed helix, right-handed helix and an intermediate configuration between the two.  To demonstrate this, we plot the free energy of the simulation as a function of the Mahalanobis distance (\eq{eq:distance}) from clusters 1 and 2 as a contour plot in \fig{fig:beaded_helix}B.  There are two distinct stable states in this 2D free energy surface, each of which is centered around a small separation from either cluster 1 or cluster 2.  These are the left- and right-handed helices. Cluster identities from the three cluster model of weighted shape-GMM are indicated by the three colors of points on the same plot. The cluster labeling does a good job distinguishing fully folded (yellow and teal) from partly folded (purple) configurations as well as distinguishing the left- and right- handed helical configurations picked out by the centers of two clusters. This result is encouraging, because it demonstrates that shape-GMM is able to identify not only quasiharmonic metastable states, but also separate out and classify together intermediate structures.

\begin{figure*}[t]
    \centering
    \includegraphics[width=\textwidth]{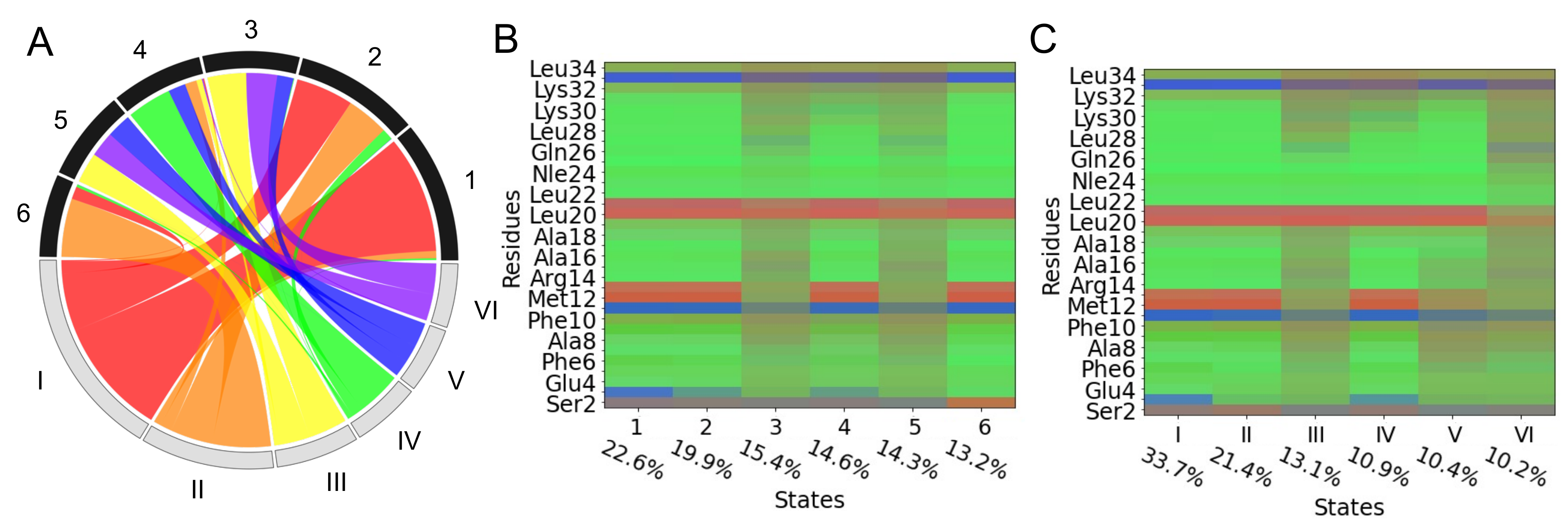}
   \caption{Uniform and weighted shape-GMM 6-state clustering results on HP35 Nle/Nle mutant. A) Matching wheel for cluster trajectories of uniform (top, 1-6) and weighted (bottom, I-VI). B) Ramacolor plot of uniform shape-GMM and C) weighted shape-GMM. Percentages correspond to cluster populations.}
  \label{fig:uniform_v_weighted_hp35}
\end{figure*}

\subsection{Application to HP35}

HP35, the C-terminal subdomain of villin, has been the focus of numerous experimental and theoretical studies of protein folding due to its ability to autonomously fold.\cite{McKnight1996,McKnight1997} The 35-residue chicken villin headpiece protein contains three helices with helix 1 composed of residues 4-10, helix 2 of residues 15-19, and helix 3 of residues 23-32. Wild-type HP35 has a folding rate of (4.3 $\pm$ 0.6 $\mu$s$)^{-1}$,\cite{Kubelka2003} which was further observed in atomistic molecular dynamics simulations with implicit solvent.\cite{Zagrovic2002} Mutations were sought to enhance the folding rate\cite{Chiu2005} and ultimately yielded a fast folding double mutant referred to as Nle/Nle (Nle - norleucine) with a measured folding rate of (0.73$\pm$ 0.05 $\mu$s$)^{-1}$ at 300K.\cite{Kubelka2006} The $\sim305\ \mu$s all-atom, explicit solvent, molecular dynamics simulation of the HP35 Nle/Nle mutant from D. E. Shaw Research has been extensively studied and serves as a benchmark system for structural clustering.\cite{Piana2012, Banushkina2013, Best2013, Sittel2014, Jain2014, Ernst2015, Mori2016,Sittel2016,Sittel2017,Nagel2019,Damjanovic2021} Here, we use this simulation at 360 K to compare and contrast between uniform and weighted shape-GMM, compare weighted shape-GMM clusterings to previous results, and, finally, propose the structural folding mechanism most consistent with the weighted shape-GMM results.

\subsubsection{Uniform vs. Weighted Shape-GMM for HP35}

We hypothesize that uniform and weighted shape-GMM will give consistent clustering results for well-folded states but inconsistent results for partially and completely unfolded states.  This hypothesis is consistent with the cluster results from the ANMs in Sec. \ref{sec:uniform_v_weighted}.  To test this hypothesis, we trained 6-state uniform and weighted shape-GMMs on $\sim$61K frames of the HP35 trajectory. For reference, a reasonable training set size should include at least $25$K frames to sample one of the observed 61 folding/unfolding events within the $\sim$1.5M-frame trajectory.\cite{Piana2012} A 6-state model was chosen due to literature precedent and simplicity of comparison (results from a scan of cluster sizes are discussed in Sec.~\ref{sec:fourstate}).  Backbone atoms (C, CA and N from residues 2 through 34, C from residue 1 and N from residue 35) are selected for the feature set to remain consistent with previous backbone dihedral value clustering methods.\cite{Jain2014,Damjanovic2021}  For each covariance approximation, 20 models with different randomly chosen starting conditions were trained and the model with the largest log likelihood was used for subsequent analysis.

Uniform and weighted 6-state shape-GMMs show consistency in separating the folded from unfolded states of HP35.  To support this claim, we combine clustering overlap between the two models as assessed by a matching wheel (\fig{fig:uniform_v_weighted_hp35}A) and stability of dihedral distributions of the residues in each cluster represented in ramacolor plots\cite{Sittel2016} (\fig{fig:uniform_v_weighted_hp35}B for uniform and \fig{fig:uniform_v_weighted_hp35}C for weighted). The backbone dihedral to color array mapping can be found in Figure S2 .  Starting with the weighted shape-GMM 6-state ramacolor plot (\fig{fig:uniform_v_weighted_hp35}C), we observe that states I and II have bright coloration for all residues indicating well-defined dihedral distributions in these states.  State IV has the majority of residues in well-defined dihedral states with the C-terminus starting to display some dihedral disorder as evidenced by the muted color for residues 30-34.  The remaining weighted shape-GMM states (III, V, and VI) have over half of the residues with muted ramacolor distributions indicating that these states are relatively disordered.  Thus we conclude that weighted shape-GMM states I, II, and IV are folded states and III, V, and VI are unfolded.  The matching wheel indicates that weighted shape-GMM folded states (I, II, and IV) are predominantly matched with states 1, 2, 4, and 6 from uniform shape-GMM.  The ramacolor plot for uniform shape-GMM indicates that states 1, 2, 4, and 6 are all fairly structured in their dihedral distributions and that the remaining clusters (3 and 5) are disordered.  Thus we conclude that uniform and weighted shape-GMM are both able to separate folded from unfolded states.

Weighted shape-GMM separates states by distinct regions of instability corresponding to secondary structure elements in HP35. Considering the most unstable uniform states 3 and 5, indicated by muted colors across nearly all residues, there are no obvious differences between these two states indicated in \fig{fig:uniform_v_weighted_hp35}B. In contrast, the unstable weighted states III, V and VI have clear structural differences (\fig{fig:uniform_v_weighted_hp35}C). The dihedral distributions of residues within weighted state III closely resemble either uniform states 3 and 5, where all three helices show instability but there is stability in turn 2 and in some subsequent residues. Alternatively, weighted state VI is entirely unstable, but gets divided up evenly to define uniform states 3 and 5. Weighted state V has stability in helix 3, and helix 2, and is composed of frames that make uniform states 3, 4, and 5. Finally, weighted state IV is a very stable state besides the C-terminal end, which compares reasonably well with the uniform state 4 besides the weighted state showing more stability in the N-terminal helix. Altogether, these results confirm that weighted shape-GMM improves structural discrimination when it comes to flexibility in particular regions, whether it's a single residue or secondary structures. Therefore, for identifying states in challenging biomolecular processes such as protein folding, a weighted shape-GMM is preferred.

The 6-state weighted shape-GMM trained on backbone atoms captures several structural distinctions between unfolded, intermediate and native states previously suggested by backbone dihedral clustering schemes for 6- and 12-state models.\cite{Jain2014,Sittel2016,Sittel2017,Damjanovic2021} For example, the Ramacolor plot for this model in \fig{fig:uniform_v_weighted_hp35}C illustrates that states I and II differ in the conformation of Asp3. The helical conformation of Asp3 has previously been shown to distinguish between native and intermediate states in 6- and 12-cluster models.\cite{Jain2014,Sittel2016} That weighted shape-GMM predicts the native state to be larger in population than the intermediate state is inconsistent with some previous results\cite{Sittel2016} but consistent with other recent results.\cite{Damjanovic2021} 
A native-like state that is structurally similar to the native state, but differing mainly in the dynamics due to an unlocked and partially unfolded helix 3 has been identified previously, and is in line with state IV from weighted shape-GMM.\cite{Reiner2010,Jain2014,Banushkina2013, Beauchamp2012,Serrano2012} Alternatively, state V indicates the opposite trend; the N-terminus and helix 1 are highly fluctuating, and there is more stability from residue 17 leading into helix 3, with a slight increase in positional variance again at the C-terminus. This indicates state V has an unfolded helix 1, but partial folding of helices 2 and 3. 
These structural attributes have also been discussed in previous models, and used to define unfolded states in more detailed 12-state models.\cite{Sittel2016} The ability of our weighted shape-GMM to capture these features in a 6-state model supports the high capacity for structural discrimination and the model's prioritization of these states as distinct.

\subsubsection{HP35 Folding: An Intuitive 4-state Model}
\label{sec:fourstate}

Weighted shape-GMM results on HP35 are most consistent with a 4-state model.  A 10-fold cross-validation scheme using a $\sim25K$ frame training set applied to a scan of cluster sizes shows a significant \% change in the slope of the log likelihood at a cluster size of 4 (see Figure S1) which suggests that 4 clusters is an appropriate choice given the structural data used to fit the model. A cluster size of 8 could also be considered with this reasoning, however the deviation of log likelihood between the training and prediction data is an indication of model overfitting. It is possible that with this large of a cluster size the training data size must be increased to build a model with representation in the training data for 8 states to be sufficiently clustered. 

\begin{figure}
    \centering
    \includegraphics[width=0.8\columnwidth]{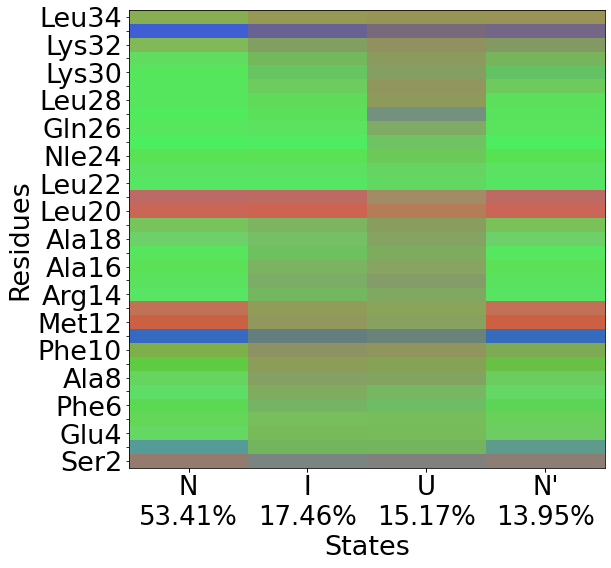}
    \caption{Ramacolor plot for the HP35 weighted shape-GMM 4-state model prior to dynamical coring.}
    \label{fig:4-state-ramacolor}
\end{figure}

The 4-state weighted shape-GMM is composed of a native state, near native state, intermediate state, and an unfolded state.  The most populated cluster (53.41\%) is characterized as the native state, N, with completely folded helices and the ${\alpha}_L$-Asp3  conformation (\fig{fig:4-state-ramacolor}). The native-like state, N', that resembles N, but differs in the partial unfolding of helix 3, is the least populated (13.95\%). In contrast to the previous 6-state model, a near-native intermediate state with completely folded helices and a ${\alpha}_R$-Asp3 is not prioritized as a distinct state with a cluster size of 4. This model still identifies an intermediate state, I (17.46\%) with an unfolded helix 1, and partial folding in helices 2 and 3, and a broadly disordered unfolded state, U (15.17\%). 

\begin{figure*}
    \centering
    \includegraphics[width=\textwidth]{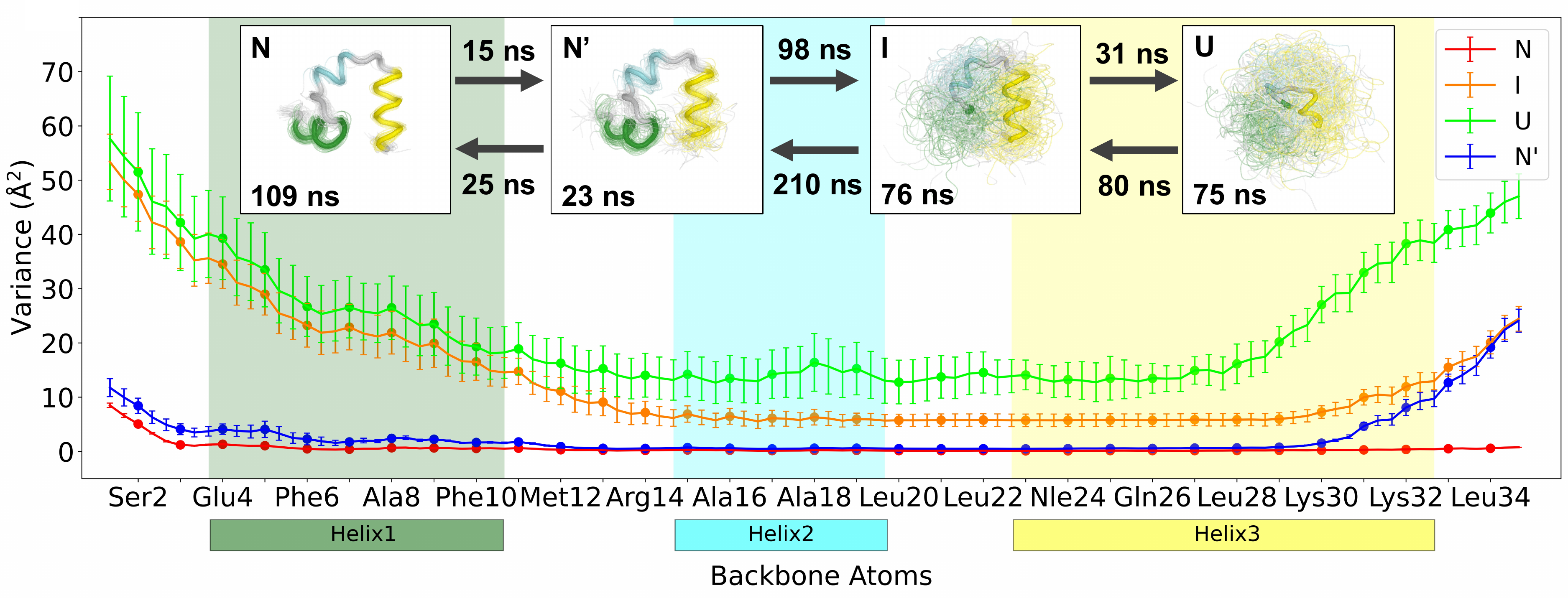}
    \caption{Weighted 4-state shape-GMM results for HP35. Variation in backbone atomic displacement of each state. Regions corresponding to helix 1, helix 2 and helix 3 are colored forest green, cyan, and yellow, respectively.  Error bars represent standard deviation of the values estimated from dividing the trajectory into 10 equal sized continuous segments.  Enlarged points represent the C$_\alpha$ atoms of each residue. Top: GMM centers in thick tube representation with 100 random frames from each cluster in thin tube representation. State lifetimes are shown in each structural figure, with MFPTs between connected states.}
    \label{fig:4-state-MSM}
\end{figure*}

A Markov state model (MSM) with a lag time of 2 ns, built from the 4-state shape-GMM results using PyEMMA\cite{scherer2015pyemma}, yields three dynamically distinct processes. The trajectory of cluster identities was first fed through a dynamic coring procedure with a minimum window of 2 ns (10 frames).\cite{Nagel2019} This changed only 3.8\% of frames suggesting that the results from weighted shape-GMM were already dynamically stable.  The 4-state model yields a passing Chapman-Kolmogrov test (see Figure S3). Three implied timescales are observed, having values of 348 ns, 40.9 ns, and 19.1 ns.  The processes associated with these three timescales are population shifts from states I and U to states N and N', between states I and U, and between states N and N', respectively. Additional results from this model are depicted in the network diagram presented in \fig{fig:4-state-MSM}. 

The folding mechanism of HP35 (Nle/Nle mutant) proceeds from the unfolded state, U, through an intermediate state, I, and near-native state, N', to the native state, N.  The mean first passage time (MFPT) to go from U to I is 80 ns and this process is characterized by a partial structuring of the C-terminal end of helix 3 (residues 29-34).  Subsequent folding of helices 1 and 2 (I to N') is the rate-limiting step, requiring 210 ns. Once the N-terminal helix has formed, the final step is to completely fold the C-terminal helix, taking 0.12 $\mu$s. Previous models proposed a native-like state that serves as a kinetic trap, only accessible through N,\cite{Sittel2014} however, our model shows N' is en route to N, which is consistent with experimental evidence.\cite{Reiner2010} The unfolding mechanism is similar to folding, but in reverse.  In both cases, the transition states are observed between the N' and I states as quantified by the committors. 

The structural characteristics of the intermediate state support a mechanism that primarily folds helix 3 and 2 before helix 1. The order of helix formation during HP35 folding has been heavily debated, both experimentally and theoretically, with some evidence that order is highly influenced by both physical and methodological factors, such as temperature, or force-field used.\cite{Mori2016,Reiner2010,Wang2019,Nagarajan2018,Jain2014,Piana2011} A recent MSM of Nle/Nle HP35, built from a 57-state model, found that over 90\% of all pathways formed helix 3 before helices 2 and 1.\cite{Nagel2019} The most preferred pathway appeared to fold in the order of helix 3, then 2, and lastly 1, which agrees well with our much more reduced model. Additionally, recent studies show evidence of Nle/Nle folding beginning with stabilization in the middle of helix 3 at residue 28, to the helix 3 N-terminus at residue 22, rather than a fully folded helix 3,\cite{Mori2016,Wang2019} which aligns well with our results that show even in the unfolded state the N-terminus of helix 3 has the lowest variance in backbone atom positions (\fig{fig:4-state-MSM}). In going from U to I, the backbone atom variances decrease significantly for residues in helices 3 and 2. Interestingly, residues 20-29 in the intermediate state appear fairly stable in that the variance is nearly the same for each backbone atom, yet the unchanging value is still high ($\sim20$ $\AA^2$). While these residues exist in stable conformations (also indicated in \fig{fig:4-state-ramacolor}), they may need to accommodate for the large fluctuations coming from the unfolded helix 1 and loop 1. This indicates that even though these regions are folded, their stability can be influenced by the rest of the protein. Furthermore, this information is a direct result of weighted shape-GMM, as the covariance is iteratively evaluated and used to optimally define each cluster.

A MSM based on our weighted shape-GMM 4-state discretization represents a respectable balance between structural resolution and kinetic detail. The mechanism of folding aligns well with previous models, even those based on much larger MSMs, and captures subtle but relevant state differences, such as the increased C-terminal variance in the native-like state. It is important to reiterate that this model was fit on $\sim4\%$ of the total trajectory data, therefore, not all folding events were represented during training. It is becoming more evident that although HP35 is a small protein, the folding mechanism is complex and likely to result from multiple pathways with possible kinetic traps and misfolded states. If the goal was to model an ensemble of folding pathways, which have been suggested for this system, then we propose: 1) using a larger training size to invite more representations for potential pathways, such that larger cluster sizes (e.g. 8) can be chosen without overfitting, and 2) exploring additional features, such as side chain center of mass, since it has been proposed that for HP35 the secondary structures form first, primarily in the unfolded basin, and that then longer scale side-chain ordering must occur to achieve native contacts.\cite{Hu2010,Best2013} As a measure of MSM validation, VAMP-2 scores for various HP35 discretizations, with and without dynamical coring are provided in Table 1 of the SI. The VAMP-2 scores of 4-state, 6-state, and 12-state Shape-GMM results compare favorably with a well-informed 12-state model following the most probable path (MPP) clustering protocol designed by Nagel et al.(\citenum{Nagel2019}), especially when the clusterings are further refined using dynamical coring. This supports the viability of this structure-based method in yielding kinetically relevant information, even though the procedure is void of explicit dynamic information.

\subsection{Practical Considerations for shape-GMM}

Shape-GMM is an appealing approach to structural clustering but is not without its limitations.  There are three major points to consider when using this method: (1) difficulty in convergence to global maxima, (2) estimating high dimensional objects, and (3) the computational expense of fitting the models.  The first of these limitations is true for any EM procedure.  This concern can be alleviated by performing each maximization numerous times with different model initialization parameters and selecting the model with the highest resulting log likelihood, as suggested by others.\cite{Do2008}  While there may be additional improvements that can be done to ensure global maxima convergence, we believe that the other two limitations are more specific to the application of shape-GMM to MD data and are discussed in more detail in the following subsections.

\subsubsection{Clustering Convergence and Estimating $N\times N$ Covariances.}

The most challenging component of estimating a weighted shape-GMM are the multiple, high dimensional, covariance matrices.  To estimate a weighted shape-GMM for $K$ clusters on a trajectory with $N$ atoms, one must estimate $K$ $N\times N$ covariance matrices.  The minimum criteria is to make each matrix full rank which requires $N+1$ independent data points for each matrix.  Shape-GMM is aided in this endeavor by two aspects of the model.  First, there are three independent measures of the $N\times N$ covariance in each frame.  Second, each frame can contribute to the covariance matrix of each cluster as indicated in \eq{update_var}.  The weight of each frame is determined by the posterior distribution (\eq{posterior}) and can be vanishingly small.  Practically speaking, one must have multiple ($>10$) independent samples of each covariance to have any confidence in their estimates.   Thus, a reasonable estimate is $\alpha \frac{(N+1)}{3}$ frames with $\alpha > 10$ for a training set.  The quality of the estimate of the covariance will be dictated by $\alpha$ and the value needed to perform clustering will depend on the system being clustered. 

\begin{figure}
    \centering
    \includegraphics[width=\columnwidth]{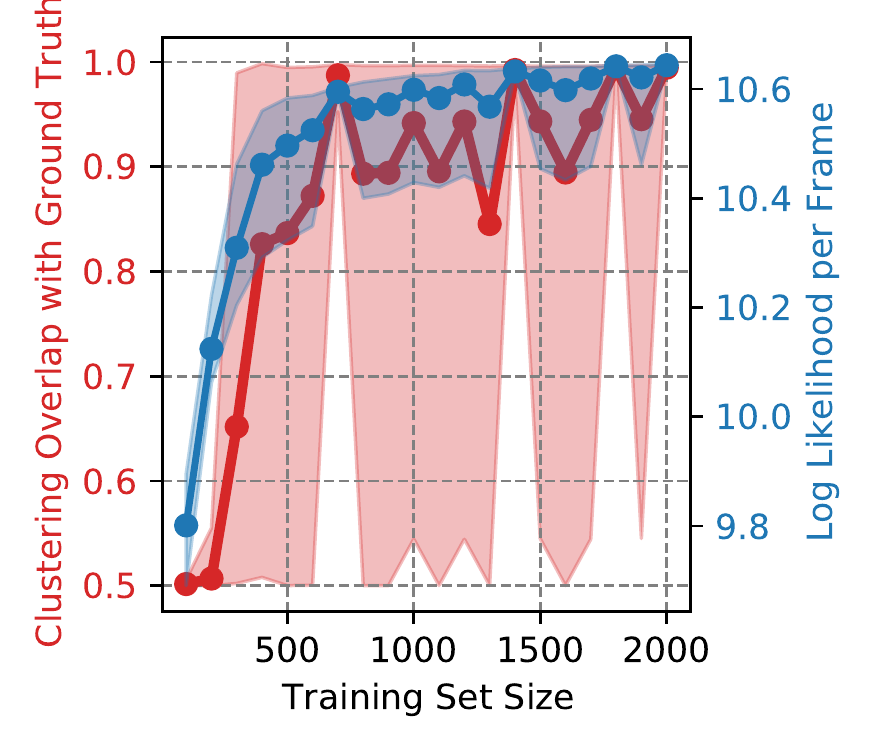}
    \caption{Clustering convergence for weighted shape-GMM on the PUBS 1 PUBS ANM system.  The average clustering overlap with ground truth and log likelihood per frame of the are plotted as a function of training set.  Each training set was chosen randomly from the 10K total frames.  This procedure was repeated 10 times to compute the average and 90\% confidence interval (shaded region).}
    \label{fig:clustering_convergence}
\end{figure}

The PUBS 1 and PUBS 2 ANM models with $n=8$ provides a challenging system to cluster because the difference in the two structures is heavily dependent on the difference in covariances.  Thus, we investigate the convergence of weighted shape-GMM on this data as function of training set size (\fig{fig:clustering_convergence}).  We consider training sets of sizes varying from 100 to 2000 frames with each training set picked randomly from the total of 10K frames.  Each training set size was sampled ten times and the resulting models were used to predict clustering on the full 10K frame trajectory.  The resulting clustering overlap with the ground truth and log likelihood per frame are depicted in \fig{fig:clustering_convergence}.  For training sets of size $< 300$ frames, we see that the clustering overlap with ground truth is low ($\approx 50$ \%) but that this overlap rapidly increases from training set sizes 300 to 700.  From there, there is some oscillation of the average value but overlaps are all near or above 90 \%.  The log likelihood per frame of the entire clustering show a similar but more steady increase and plateauing behavior.  We note that the clustering overlap for models trained on even $>1000$ frames still shows significant deviation between 50 \% and 100 \% overlap with ground truth as indicated by the shaded region.  This indicates the importance of the particular training set chosen, especially for relatively small training sets. Therefore, it is recommended to fit shape-GMM on various training sets and use the model with the highest log likelihood on the entire data set.  

As $N$ gets large, sampling of the covariance matrices will become more challenging.  It has previously been estimated that RMSD-based clustering will start to fail for greater than 200 particles.\cite{Sargsyan2017}  This analysis utilized both a standard RMSD measure as well as a weighted RMSD measure, but this weighted measure still equates to a diagonal covariance matrix, not the weighted form of the covariance we describe in this paper.  Thus, it remains an open question as to the limit of $N$ for our weighted shape-GMM clustering.  All systems studied in this paper fall under the 200 particle threshold given for RMSD.

\subsection{Computational Expense of Shape-GMM}

\begin{figure}
    \centering
    \includegraphics[width=\columnwidth]{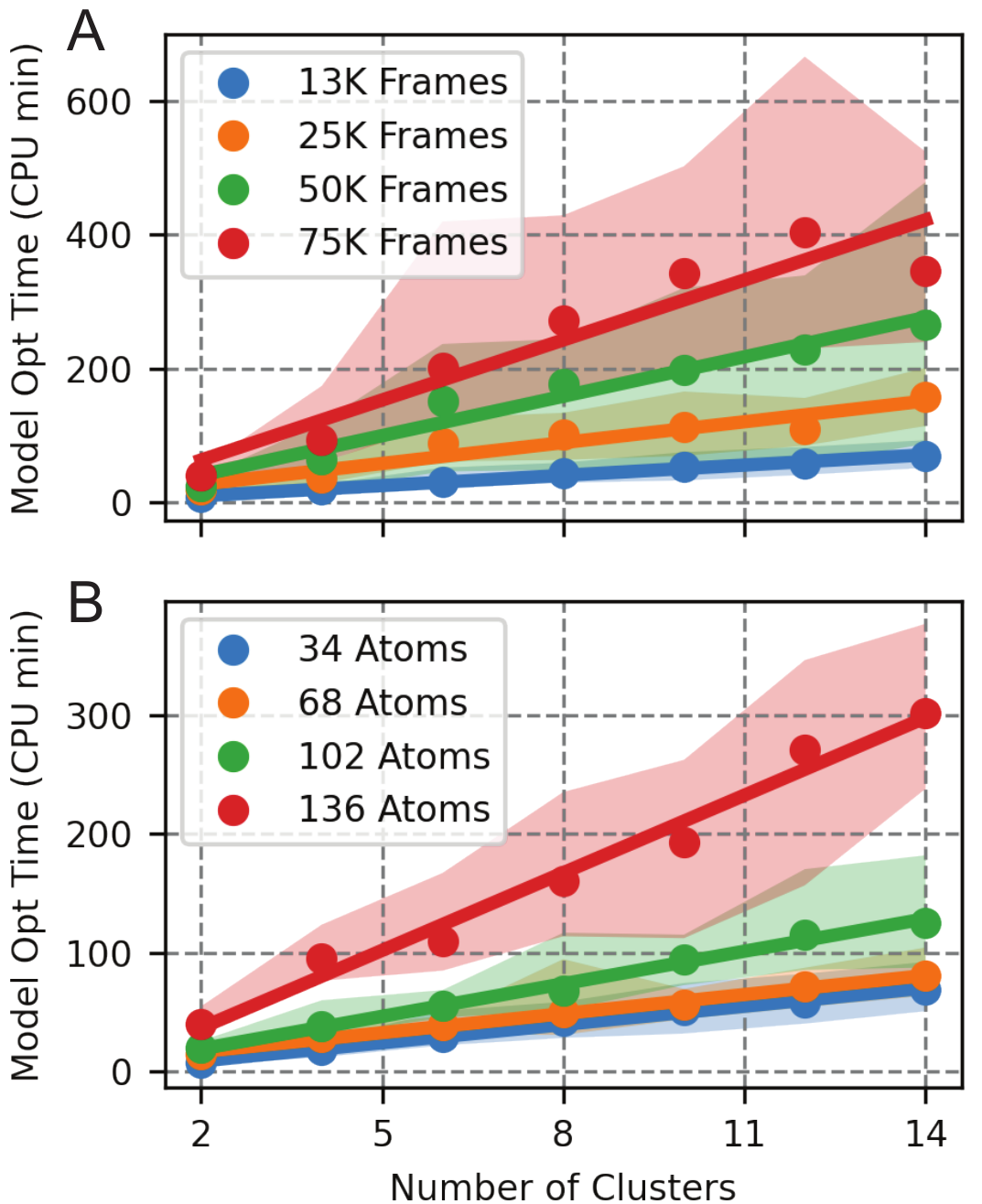}
    \caption{CPU time for weighted shape-GMM optimization for HP35. A) CPU time to optimize a single weighted shape-GMM on 34 atoms as a function of number of clusters for different training set sizes. B) CPU time to optimize a single weighted shape-GMM on $\sim$ 13K frames as a function of number of clusters for different number of atoms in the feature space.}
    \label{fig:timescaling}
\end{figure}

Computational expense of weighted shape-GMM scales linearly with the number of clusters.  This can be observed in both plots depicted in \fig{fig:timescaling}.  We focus on the orange curve with a training set of $25$K frames in \fig{fig:timescaling}A as this is the same training set size used for our cluster scan.  We observe that for 2 clusters, a weighted shape-GMM takes approximately 10 minutes to optimize.  For 14 clusters, it takes approximately 150 minutes to optimize.  The large variance in optimization time, as indicated by 90\% confidence intervals surrounding the solid lines, is a natural aspect of the EM procedure for GMMs.  The time it takes to find a maximum in log likelihood depends dramatically on the starting conditions.  Regardless, the trend in the average time as a function of number of clusters is clearly linear, as one would expect for a serial implementation of the EM GMM procedure.  

Computational expense scales linearly with the number of frames in the training set.  The four different lines in \fig{fig:timescaling}A indicate different training set sizes.  We start with $12.6$K frames and then increase by multiples of $2$ to include $\sim25$K, $50$K, and $75$K frames.  The slopes of the best fit lines are $5.1$, $10.4$, $19.7$, and $30.0$ CPU min/cluster, respectively.  These scale by the same multiplicitive factor as the number of frames thus indicating that the procedure scales linearly with number of frames.  

Computational expense scales quadratically with the number of atoms, beyond an initial sub-quadratic region.  The CPU time as a function of number of clusters is plotted for four different feature space sizes in \fig{fig:timescaling}B.  All of these were done with $12.6$K frames in the training set.  The behavior between different feature space sizes is not as clear as that of training set size.  The slopes of the best fit lines are $5.1$ CPU min/cluster for 34 atoms, $5.3$ CPU min/cluster for 68 atoms, $9.1$ CPU min/cluster for 102 atoms. and $21.8$ CPU min/cluster for 136 atoms.  The ratio of the slopes of the smallest feature space size (34 atoms) to that of the largest feature space size (136 atoms) is approximately 4, comparable to the ratio of the feature space sizes themselves.  The two medium-sized feature spaces behave sublinearly compared to the smallest feature space size.  Comparing the slopes of the CPU times of 68 and 136 atoms, we observe a quadratic relationship between time and feature space size.  We expect CPU time to scale as N$^2$ due to the two dominant processes being calculations of covariances.

It is possible to train weighted shape-GMMs with 100s of atoms on 10s of thousands of frames within a few hours.  The computational time is compounded by the recommendation to run multiple iterations to increases chances converging to a global maximum, as well as the potential need to scan number of clusters.  That said, each of those runs is completely independent and thus can be run simultaneously on separate processors.  Additionally, the clustering algorithm requires very few user parameters: feature space size, training set size, and number of clusters.  This is in stark contrast to many other clustering protocols.  

\section{Conclusion and outlook}

A Gaussian Mixture Model on particle positions is a conceptually appealing approach to model the high dimensional probability density of macromolecules.  The difficulty in doing so has been to account for the ability of the molecule to freely rotate and translate in a way that properly accounts for particle-particle correlation.  Here, we present the maximum likelihood alignment and GMM procedures necessary to determine optimal parameters for a given mixture size.  This procedure can be used to discretize conformational space in a way that theoretically matches the intuition of molecules hopping between harmonic free energy minima.  

Two shape-GMMs are presented: uniform, a model in which the particle covariance of each mixture is presumed to be proportional to the identity matrix, and weighted, a model in which the particle covariance of each mixture is presumed to have the form $\bm{\Sigma}=\bm{\Sigma}_N\otimes\bm{I}_3$.  Both uniform and weighted models are able to distinguish between five structurally distinct elastic network models.  

Weighted shape-GMM is able to distinguish between distinct structures for species with heterogeneous particle variation.  This finding disproves a previously held belief that particle positions cannot distinguish between these types of structures, and demonstrates an achievement of shape-GMM that will greatly benefit the field.  

Weighted shape-GMM also provides globally distinct clusters for the folding/unfolding of HP35 from an all-atom molecular dynamics trajectory.  The identified clusters corroborate some previously identified features of the conformational ensemble of HP35, but are distinct in the global picture of the folding pathway.  Specifically, a four-state model is most consistent with our data and predicts that the native state comprises 50+\% of the entire trajectory and follows a C-terminal unfolding and then N-terminal unfolding pathway.  

Weighted shape-GMM can also be used to compare clusterings from other, potentially faster procedures.  The log likelihood of a particular clustering can be readily computed and compared between clusterings.  The clustering with the largest log likelihood represents the best partitioning of the data under the given GMM.  

The generality of our approach is strengthened by the ease of access and application.  The package can be easily installed from PyPI (pip install shapeGMM) or directly from github (https://github.com/mccullaghlab/GMM-Positions).  The code interface is designed to mimic the Scikit-learn GaussianMixture package with and object initialization, fitting (albeit with uniform and weighted versions), and prediction routines.

The shape-GMM algorithm is more computationally demanding than some other clustering procedures.  While linear scaling is observed, as expected, as a function of number of clusters and training set size, quadratic scaling is observed for feature space size.  This will limit the system sizes that weighted shape-GMM can be readily applied to.  We note, however, that there are some distinct advantages, other than conceptual ones, of the algorithm: few parameters are needed for clustering, and the method can be directly applied to any type of molecular system.  Looking forward, we expect a GPU implementation of the algorithm to greatly improve the applicability for larger system sizes or data sets.

\section{Simulation Details}
\label{sec:sim_details}
\subsection{Elastic Network Model Simulations}

Elastic Network Models (ENMs) were simulated in the LAMMPS package \cite{Plimpton1995}.  Harmonic bonds were placed between each bead.  Langevin dynamics simulations were performed at 300 K in the NVT ensemble with a damping coefficient of 10 fs$^{-1}$.  An integration timestep of 2 fs was used.  Simulations were run for 10 million steps with frames written every 1000 steps.  

\subsection{Beaded Helix Simulations}
A 12-bead model designed to have two equienergetic ground states as left- and right-handed helices \cite{Hartmann2020} was simulated in LAMMPS \cite{Plimpton1995}. 
11 harmonic bonds between beads having rest length length 1.0 and spring constant 100 form a polymer backbone. Lennard-Jones (LJ) interactions between every $i,i+4$ pair of beads with $\epsilon=6.0$ and $\sigma=1.5$ and a cutoff length of 3.0 give rise to the helical shape. All non-bonded $i,i+2$ and farther also have a repulsive WCA interaction with $\epsilon=3.0$ and $\sigma=3.0$ added to prevent overlap, with the $\epsilon$ for $i,i+2$ reduced by 50\%. Simulations at temperature 1.0 were performed using `fix nvt' using a simulation timestep of 0.005 and a thermostat timestep of 0.5. 
A folding/unfolding trajectory of length 50000000 steps was generated and analyzed as above. Here, all parameters are in reduced (LJ) units. 

\subsection{HP35 Simulation}
A 305 $\mu$s all-atom simulation of Nle/Nle HP35 at 360 K from Piana et al.\cite{Piana2011} was analyzed. The simulation was performed using the Amber ff99SB*-ILDN force field and TIP3P explicit water model.
Protein configurations were saved every 200 ps, resulting in $\sim$1.5M frames.

\begin{acknowledgement}
The authors would like to thank Peter McCullagh for useful discussions on maximum likelihood procedures and size-and-shape space.
We would also like to thank D.E. Shaw Research for providing simulation data on the HP35 protein. 
GMH was supported by the National Institutes of Health through the award R35GM138312.
MM would like to acknowledge funding from National Institute of Allergy and Infectious Diseases of the National Institutes of Health under award number R01AI166050.
\end{acknowledgement}

\bibliography{refs}

\end{document}


\renewcommand{\thefigure}{S\arabic{figure}}
\renewcommand{\thetable}{S\arabic{table}}
\begin{table}[]
    \label{tab:my_label}
    \centering
    \caption{Scores for a variety of clusterings of the HP35 trajectory. VAMP-2 scores were computed after construction of MSMs using a 10-fold cross validation on five trajectory segments using PyEMMA.\cite{scherer2015pyemma}  These were computed using either the first four ($k=4$) or the first 10 ($k=10$) eigenvalues of the transition matrix.  Log likelihoods of a weighted shape-GMM (wSGMM) model were also computed on 10 randomly selected trajectory segments.  A dynamic coring (here denoted dCore) procedure was applied to each clustering\cite{Nagel2019}.  Error as estimated by 10-fold trajectory chunking are reported in parentheses.}
    \begin{tabular}{l|d{5.2}|d{5.2}|d{5.2}}
    \hline
    Clustering Model & \multicolumn{1}{p{30mm} |}{VAMP-2 Score (k=4)} & \multicolumn{1}{p{30mm} |}{VAMP-2 Score (k=10)} & \multicolumn{1}{p{30mm}}{wSGMM Log Likelihood per Frame} \\
    \hline
    wSGMM 4-state                          &  3.3(2)  &  & 372.4(1)\\
    wSGMM 4-state (dCore)                  &  3.87(3) &  & 372.0(1)\\
    wSGMM 6-state                          &  3.4(1)  &  & 379.5(1) \\
    wSGMM 6-state (dCore)                  &  3.89(1) &  & 378.6(1) \\
    uSGMM 6-state                          &  2.9(1)  &  & 368.0(1) \\
    uSGMM 6-state (dCore)                  &  3.93(1) &  & 367.66(9)\\
    wSGMM 12-state                         &  3.7(1)  &  7.2(6) & 388.4(1) \\
    wSGMM 12-state (dCore)$^*$             &  3.91(1) & 9.39(5) & 383.16(9)\\
    Stock 12-state  (dCore)                &  3.91(1) & 9.30(4) & 364.2(3)\\
    Stock 12-state  (state specific dCore)$^\dagger$ &  3.93(1) & 9.46(3) & 363.9(2)\\
    \hline
    \end{tabular}
     \vspace{1ex}
     {\raggedright $^*$ Dynamic coring of wSGMM 12-state yielded a 10-state model. \par}
\end{table}

\begin{figure}
    \centering
    \includegraphics[width=\textwidth]{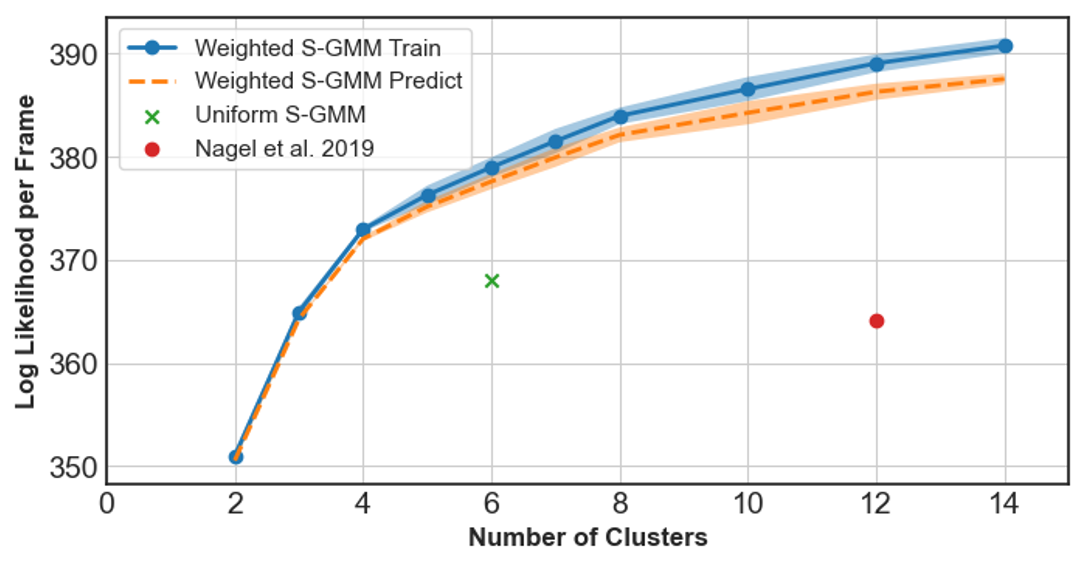}
    \caption{Log likelihood of weighted shape-GMM (S-GMM) as a function of number of clusters for HP35 trained on $\sim25K$ frames (blue line) in a 10-fold cross validation (orange dashed) scheme. The shaded region indicates the 90\% confidence interval. The features are C, CA and N backbone atoms of residues 2-34, the C atom of residue 1 and N atom of residue 35 (101 atoms total). The log likelihood value from a uniform 6-state shape-GMM is shown as a green x marker. The red circle marker indicates the log likelihood that results from the HP35 discretization of Nagel \textit{et al}. clustering procedure resulting in 12-states\cite{Nagel2019}.}
    \label{fig:hp35_log_lik}
\end{figure}
\begin{figure}
    \centering
    \includegraphics[width=0.6\textwidth]{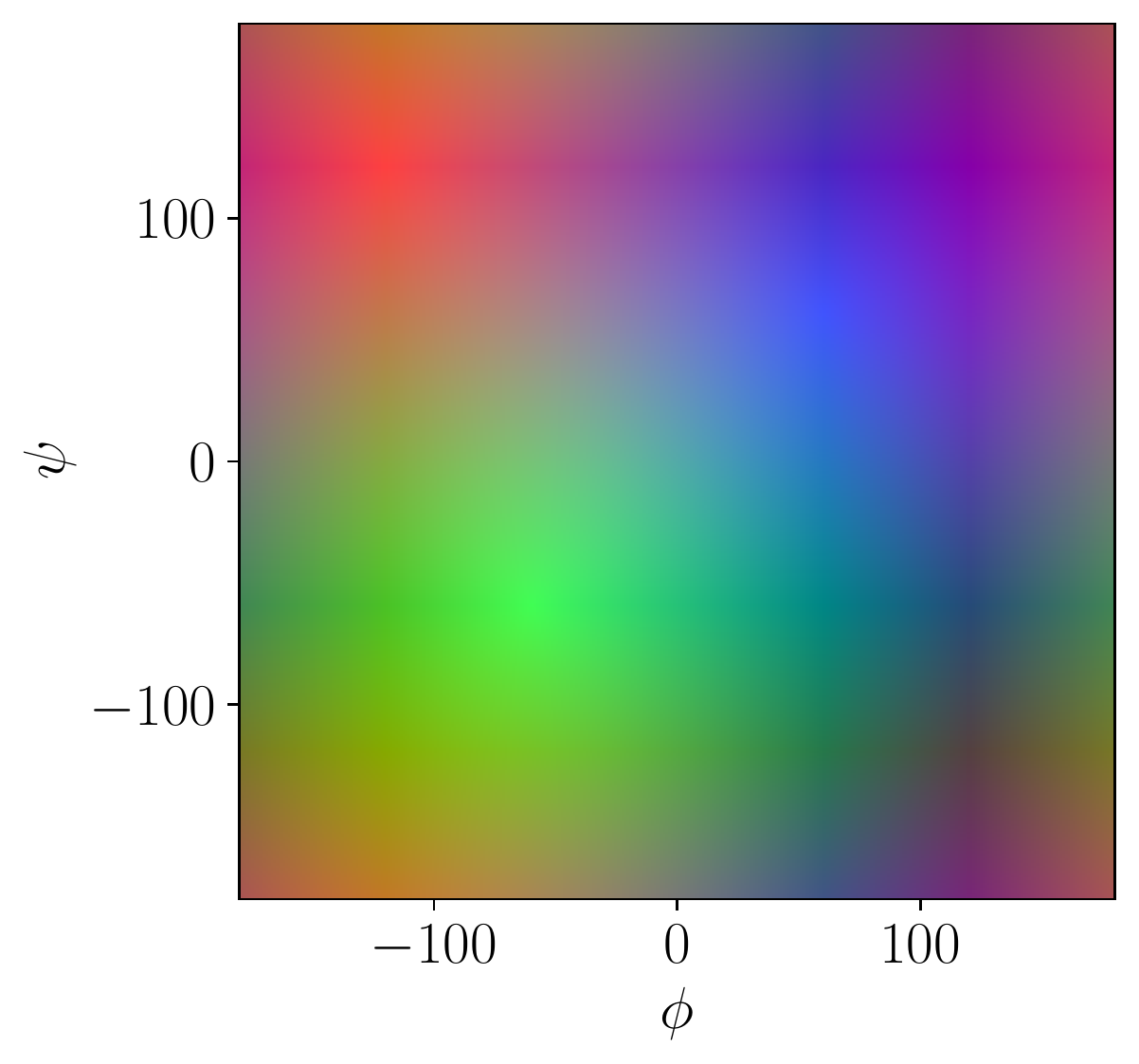}
    \caption{RGB values corresponding to phi-psi angle pairs used to assignment a unique color value to a dihedral conformation. The major dihedral states are indicated with more vibrant colors: red, green, and blue correspond respectively to $\beta$-strand, $\alpha_R$-helical, and $\alpha_L$-helical dihedral character. The code used to produce this mapping can be found at https://github.com/moldyn/ramacolor.\cite{Sittel2016,Nagel2019}}
    \label{fig:ramacolor}
\end{figure}
\begin{figure}
    \centering
    \includegraphics[width=0.9\textwidth]{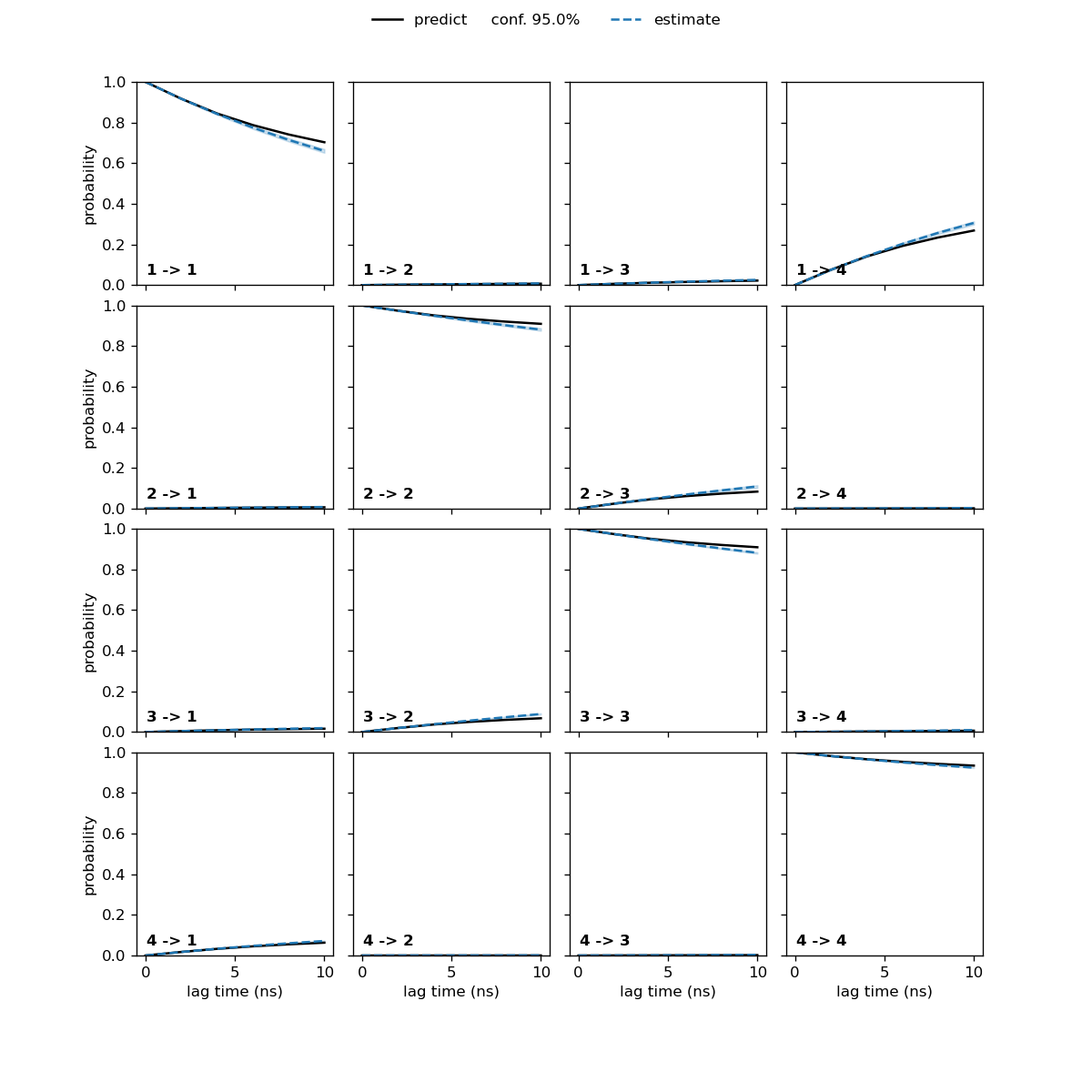}
    \caption{Chapman-Kolmogorov test for 4-state MSM resulting from weighted shape-GMM clustering on HP35.  Computed using PyEMMA software.\cite{scherer2015pyemma}}
    \label{fig:hp35_4_state_ck_test}
\end{figure}

\newpage
\bibliography{refs}